\documentclass[twocolumn,prb,preprintnumbers,amsmath,amssymb,superscriptaddress]{revtex4-1}
\usepackage[dvipdfmx]{graphicx}
\usepackage{color}
\usepackage{dcolumn}
\usepackage{bm}
\usepackage{amsmath,amssymb}
\usepackage{ulem}

\begin{document}

\title{Distinct variation of electronic states due to annealing in La$_{1.8}$Eu$_{0.2}$CuO$_4$ and Nd$_{2}$CuO$_4$
}

\author{Shun Asano}
\affiliation{Department of Physics, Tohoku University, Aoba, Sendai 980-8578, Japan}
\affiliation{Institute for Materials Research, Tohoku University, Katahira, Sendai 980-8577, Japan}
\author{Kenji Ishii}
\affiliation{Synchrotron Radiation Research Center, National Institutes for Quantum and Radiological Science and Technology, Hyogo 679-5148, Japan}
\author{Daiju Matsumura}
\author{Takuya Tsuji}
\affiliation{Materials Sciences Research Center, Japan Atomic Energy Agency, Hyogo 679-5148, Japan}
\author{Kota Kudo}
\affiliation{Department of Physics, Tohoku University, Aoba, Sendai 980-8578, Japan}
\author{Takanori Taniguchi}
\affiliation{Institute for Materials Research, Tohoku University, Katahira, Sendai 980-8577, Japan}
\author{Shin Saito}
\author{Toshiki Sunohara}
\author{Takayuki Kawamata}
\author{Yoji Koike}
\affiliation{Department of Applied Physics, Graduate School of Engineering, Tohoku University, Sendai 980-8579, Japan}
\author{Masaki Fujita}
\email{fujita@imr.tohoku.ac.jp}
\affiliation{Institute for Materials Research, Tohoku University, Katahira, Sendai 980-8577, Japan}

\begin{abstract}

We performed Cu {\it K}-edge X-ray absorption fine structure measurements on T'-type La$_{1.8}$Eu$_{0.2}$CuO$_4$ (LECO) and Nd$_2$CuO$_4$ (NCO) to investigate the variation in the electronic state associated with the emergence of superconductivity due to annealing. The X-ray absorption near-edge structure spectra of as-sintered (AS) LECO are quite similar to those of AS NCO, indicating that the ground state of AS T'-type LECO is a Mott insulator. We found a significant variation of the electronic state at the Cu sites in LECO due to annealing. The electron density after annealing ($n_{\rm AN}$) was evaluated for both superconducting LECO and non-superconducting NCO and found to be to be 0.40 and 0.05 electrons per Cu, respectively. In LECO but not in NCO, extended X-ray absorption fine structure analysis revealed a softening in the strength of the Cu-O bond in the CuO$_2$ plane due to annealing, which is consistent with the screening effect on phonons in the metallic state. Since the amounts of oxygen loss due to annealing ($\delta$) for LECO and NCO are comparable with each other, these results suggest distinct electron-doping processes in the two compounds. That electron-doping in NCO approximately follows the relation $n_{\rm AN}=2\delta$ can be understood if electrons are doped through oxygen deficiency, but the anneal-induced metallic nature and large $n_{\rm AN}$ of LECO suggest a variation of the electronic band structure causes self-doping of carriers. The origin of the difference in doping processes due to annealing is discussed in connection with the size of the charge transfer gap. 

\end{abstract}


\maketitle


\section{Introduction}

Single-layer cuprate oxides $RE_2$CuO$_4$ (where $RE$ means rare-earth) possess three structural isomers. The T'-type isomer has \color{black} four oxygens coordinated around the Cu ion, the T*-type has five, and the T-type has six. Historically, all forms of $RE_2$CuO$_4$, regardless of their oxygen-coordination numbers, have been assumed to be charge transfer (CT) Mott insulators, and the parent compounds of high-transition-temperature superconductors. However, the reported observation of superconductivity in T'-type $RE_2$CuO$_4$ has raised the possibility that the ground states of $RE_2$CuO$_4$ structural isomers are not unique ~\cite{Matsumoto2009, Asai2011, Takamatsu2012}; this inaugurated the study of how different oxygen coordinations affect physical properties of the isomers~\cite{Das2009, Weber2010a, Weber2010}. 

So far, superconductivity in Ce-free (undoped) T'-type $RE_2$CuO$_4$ has been observed in limited samples such as thin films after adequate oxygen reduction annealing\cite{Matsumoto2009}. In the case of single-crystal and high-temperature synthesized powder samples of $RE_{2-x}$Ce$_x$CuO$_4$, grown by the conventional method, superconductivity has not been realized at $x$ = 0 even for annealed samples\cite{Takagi1989,Brinkmann1995,Fujita2003,Horio2015a}. Given that the ground state of the T'-type $ RE_2$CuO$_4 $ is metallic, the removal of partially existing excess oxygens at the apical site is the key to achieving superconductivity; this can be done homogeneously for the thin film because of the large surface-to-volume ratio. 

Recent studies of photoemission and soft x-ray absorption spectroscopy have shown evidence of anneal-induced electron-doping\cite{Horio2015a, Wei2016, Song2017, Horio2018e, Horio2018b, Horio2018f}. 
On the basis of angle-resolved photoemission spectroscopy experiments on thin-film Pr$_2$CuO$_4$, Horio et al. have concluded that oxygen non-stoichiometry induced by annealing in the CuO$_2$ plane and/or the $RE_2$O$_2$ layer is the origin of such doping\cite{Horio2018e}. 
Importantly, this suggests that structural disorder is present even in the superconducting sample and that the superconductivity in T'-type $RE_2$CuO$_4$ is the result of electron doping. Thus, it is essential to understand the relationship between the anneal-induced variation of the electronic state and the occupancy (or lack of occupancy) of each site by oxygens. 

There are two possible scenarios for electron doping through annealing: 

(i) Electron doping into the Cu 3$d_ {x^2-y^2} $ upper Hubbard band (UHB) by induced oxygen deficiency and/or by removal of the apical oxygen. (See Fig. \ref{CT_Cu-Obond_v5}(c).) 
In this case, the increased electron density after annealing ($n_{\rm AN}$) is given by $n_{\rm AN} = 2\delta$, where $\delta$ is the amount of reduced oxygen in the formula unit due to annealing. 

(ii) Self-doping associated with the collapse of the CT gap\cite{Yokoyama2006, Adachi2013, Yamazaki2017}. (See Fig. \ref{CT_Cu-Obond_v5}(d).) 
It has been proposed that the removal of the apical oxygen lowers the Madelung energy of the Cu 3$d_{x^2-y^2}$ UHB\cite{Adachi2013}, resulting in the hybridization with the O 2$p$ band on the Fermi level. In this situation, both electron and hole carriers could be generated simultaneously. 

A recent X-ray absorption near-edge structure (XANES) study on high-temperature synthesized Pr$_{2-x}$Ce$_x$CuO$_4$ (PCCO)\cite{Asano2018} found that, although $n_{\rm AN} = 2\delta$ in the smaller $\delta$ region, it tends to be higher in the larger $\delta$ region. This result suggests that the sequential electron-doping process (i) gives way to the self-doping process (ii) with increasing $\delta$. 
To understand the mechanism of undoped superconductivity in T'-type $RE_2$CuO$_4$, it is necessary to clarify how annealing induces electron transfers not seen in the non-superconducting compound. For this purpose, a low-temperature synthesized powder sample of T'-type La$_{1.8}$Eu$_{0.2}$CuO$_4$ (LECO) would be an excellent candidate system\cite{Takamatsu2012}. However, superconducting LECO can be obtained only in the form of powder, and therefore, the experimental method of the electronic state is limited. 

In this study, we performed Cu {\it K}-edge X-ray absorption fine structure (XAFS) measurements on LECO and high-temperature synthesized Nd$_2$CuO$_4$ (NCO). The latter is non-superconducting even after annealing. By investigating the XANES spectra of LECO, we found significant evolution of the electronic state at Cu sites due to annealing, with a large $n_{\rm AN}$ value of 0.40 electrons per Cu. Extended X-ray absorption fine structure (EXAFS) analysis, furthermore, indicated that the strength of the Cu-O bond in the CuO$_2$ plane had softened due to annealing, which is consistent with the screening effect on phonons in the metallic state. 
In NCO, on the other hand, $n_{\rm AN}$ and $\delta$ were found to be 0.05 electrons per Cu 0.035 per unit formula, respectively, approximately following the $n_{\rm AN}=2\delta$ relation; no evidence of the softening of the Cu-O bond due to annealing was observed. Since the $\delta$ values of LECO and NCO are comparable, these results suggest that the variation of the electronic state due to annealing proceeds in a different manner in each compound.

\section{Sample preparation and XAFS experiment}

	\begin{table}[tb]
 	\begin{center}
    \caption{Lattice constants for as-sintered (AS) and annealed (AN) compounds of La$_{1.8}$Eu$_{0.2}$CuO$_4$ (LECO) and Nd$_{2}$CuO$_4$ (NCO).}
    \begin{tabular}{cccc} \hline
      ~~~~~~~~~~ & ~~~~~~~~ & ~$a$ (\AA)~ & ~$c$ (\AA)~ \\ \hline
      LECO & AS & 3.9994(7) & 12.485(4)  \\ 
      LECO & AN & 4.0037(2) & 12.459(1)\\ 
       NCO & AS & 3.9452(5) & 12.176(1) \\ 
       NCO & AN & 3.9463(5) & 12.172(1) \\ \hline
    \end{tabular}
    \label{lattice_const}
  	\end{center}
	\end{table}	

As-sintered (AS) polycrystalline samples of LECO were prepared by the low-temperature synthesis method, described previously \cite{Takamatsu2012}. Superconducting LECO with a transition temperature ($T_{\rm c}$) of 20 K was obtained by annealing the AS samples in vacuum at 700 $^{\circ}$C for 24 h. 
The oxygen loss $\delta$ in LECO due to annealing was not determined precisely but a maximum value of 0.05 per unit formula was obtained from neutron diffraction measurements~\cite{Sato}. AS polycrystalline samples of NCO were synthesized by the solid-state reaction method. 
Annealed (AN) NCO was prepared by annealing the AS NCO in flowing Ar gas at 750 $^{\circ}$C for 12 h. The value of $\delta$ in NCO was determined (by the weight loss of the sample through annealing) to be 0.035 per unit formula. AN NCO is an insulator that shows magnetic order below $\sim$260 K \cite{Suzuki2019}. 
The phase purity of the samples was checked by X-ray powder diffraction. 
Lattice constants evaluated by Rietveld analysis on the X-ray diffraction pattern for LECO and NCO are shown in Table \ref{lattice_const}. In both LECO and NCO, the in-plane (out-of-plane) lattice slightly elongates (shrinks) due to annealing. 

	\begin{figure}[tb]
	\begin{center}
	\includegraphics[width=82mm]{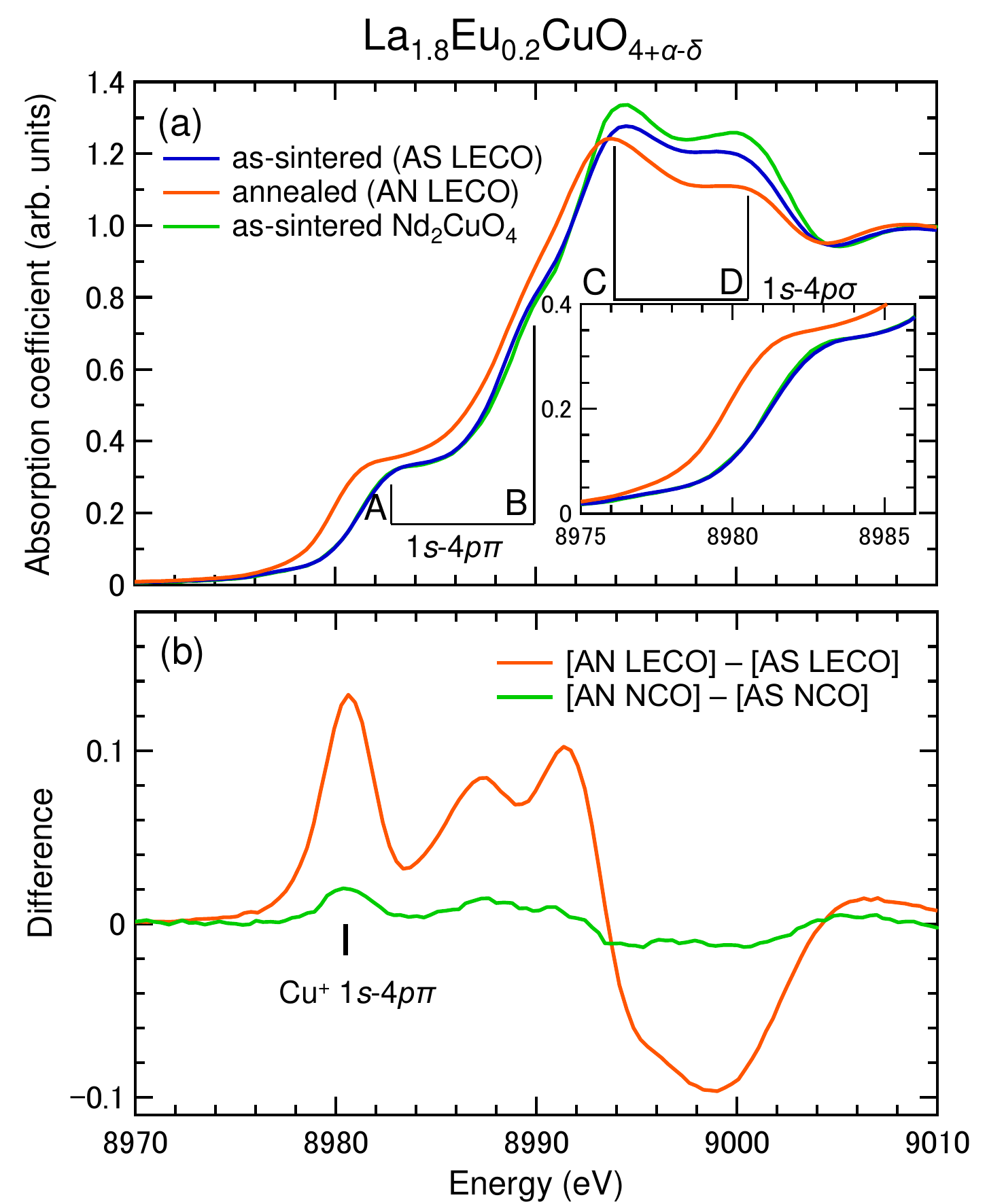}
	\caption{(Color online) (a) Cu {\it K}-edge XANES spectra for as-sintered (AS)  and annealed (AN) La$_{1.8}$Eu$_{0.2}$CuO$_4$ (LECO). The XANES spectrum for as-sintered Nd$_2$CuO$_4$ (NCO) is plotted as a reference. The inset is the enlarged spectra for the energy between 8975~eV and 8986~eV. (b) The difference spectra for LECO and NCO, which are obtained by subtracting the spectra of the AS compounds from those of AN ones. }
	\label{LECO_XANES_v3}
	\end{center}
	\end{figure}
	
Cu $K$-edge XAFS measurements were performed with the transmission mode at the BL01B1 and BL14B1 beamlines at the SPring-8 synchrotron radiation facility.
Using an Si(111) double-crystal monochromator, we measured XAFS spectra at 300~K on small pellets (7 mm in diameter and 0.5 mm in thickness), which were mixed with boron nitride for self-support.  
XAFS spectra consist of XANES and EXAFS spectra, reflecting the unoccupied electronic state and the local structure around the Cu site, respectively. The temperature dependence of the EXAFS spectra was measured from 10 K to 300 K to analyze the atomic displacement factor $C{_2}(=\sigma_{\rm s}+\sigma_{\rm d})$ of the Cu-O$_{\rm p}$ bond. Here, $\sigma_{\rm s}$ ($\sigma_{\rm d}$) is the static (dynamical) displacement component and is attributed to the random displacement of atomic positions (thermal vibration of atoms). O$_{\rm p}$ represents oxygen in the CuO$_2$ plane.

\section{Results}

	\begin{figure}[tb]
	\begin{center}
	\includegraphics[width=\linewidth]{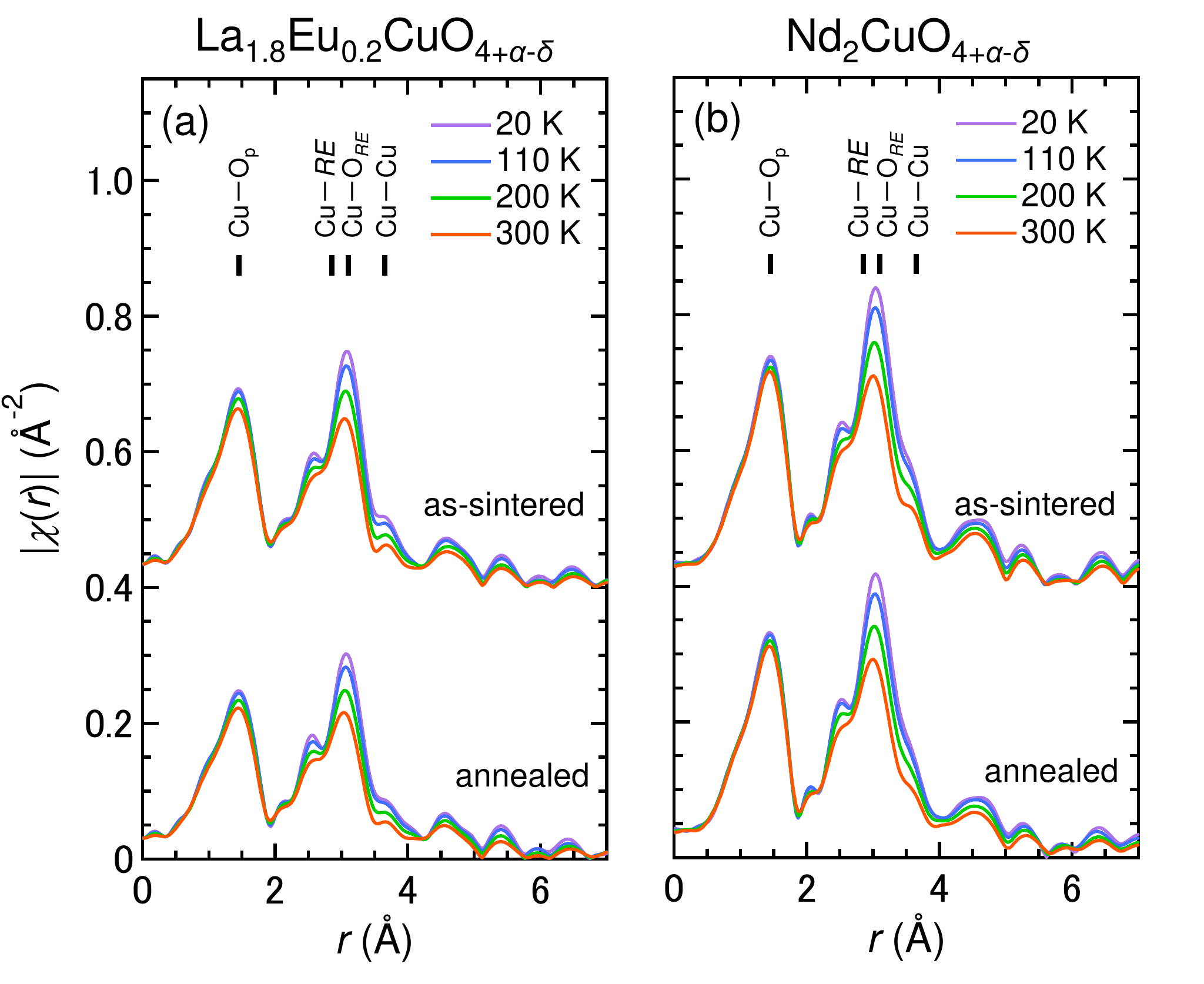}
	\caption{(Color online) The absolute value of the Fourier transform of the EXAFS oscillations for (a) La$_{1.8}$Eu$_{0.2}$CuO$_4$ and  (b) Nd$_{2}$CuO$_4$ measured at 20, 110, 200, and 300 K. The solid bars denote the corresponding positions for Cu--O$_{\rm p}$, Cu--$RE$ ($RE$ = La, Eu, Nd), Cu-O$_{\rm RE}$, and Cu-Cu paths. O$_{\rm p}$ and O$_{\rm RE}$ represent oxygen sites in the CuO$_2$ plane and in the {\it RE}O layer, respectively. The results for AS samples are shifted along the vertical direction by 0.4 ${\rm \AA}^{-2}$.}
	\label{LECO_ChiR}
	\end{center}
	\end{figure}

Figure \ref{LECO_XANES_v3}(a) shows XANES spectra for AS and AN LECO. The spectrum for AS NCO is also plotted in Fig. \ref{LECO_XANES_v3}(b) as a reference. The AS samples of LECO and NCO exhibit quite similar spectra, indicating the same electronic state in both AS compounds. This similarity is consistent with the published results for AS Pr$_2$CuO$_4$ \cite{Asano2018}. 
The small difference in the structure around 8994 eV and 9000 eV, energies corresponding to 1$s$-4$p\sigma$ transitions, can be understood by the variation of the size of rare-earth atoms\cite{Liang1995}. Therefore, the ground state of AS T'-type $RE_2$CuO$_4$ does not depend on the synthesis method. The identification of the same ground state in AS LECO and NCO is an important starting point for discussing the variation of the electronic state in the two compounds due to annealing. 

As seen in Fig. \ref{LECO_XANES_v3}(a), annealing induces a drastic change in the XANES spectrum of LECO.  The intensity around energies corresponding to the 1{\it s}--4{\it p}$\pi$ transitions (8983 eV and 8991 eV) increases, while that around the 1{\it s}--4{\it p}$\sigma$ transitions (8994 eV and 9000 eV) decreases. A similar spectral change has been observed for PCCO with Ce-substitution; it suggests electron doping into the sample\citep{Oyanagi1990, Kosugi1990, Liang1995, Asano2018}. 

	\begin{figure}[tb]
	\begin{center}
	\includegraphics[width=\linewidth]{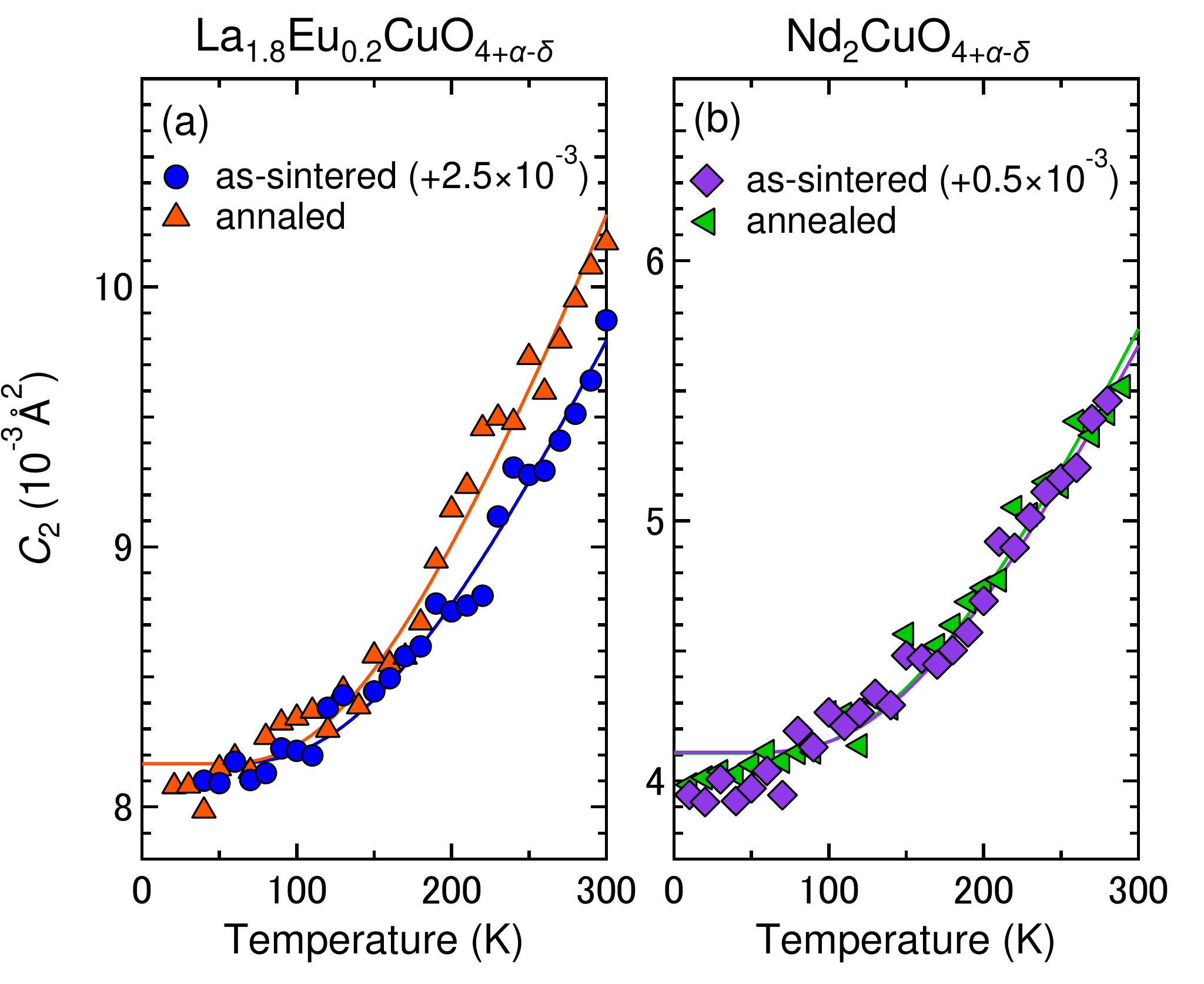}
	\caption{(Color online) Temperature dependence of the atomic displacement factor for the Cu-O$_{\rm p}$ bond $C_2$ in  (a) as-sintered and annealed La$_{1.8}$Eu$_{0.2}$CuO$_4$ and (b) Nd$_2$CuO$_4$. Solid lines are the fitting results using Eq. (\ref{eq:debye_waller}).}
	\label{DebyeWaller_v6}
	\end{center}
	\end{figure}

To see the spectral change due to annealing more clearly, we subtract the XANES spectrum of AS LECO from that of AN LECO. 
The difference spectrum of LECO is shown in Fig. \ref{LECO_XANES_v3}(b) together with the analogous result for NCO. The amplitude of the difference spectrum is much larger in LECO, demonstrating that annealing has a larger effect on the electronic state. The peak at 8981 eV in the difference spectrum reflects the 1$s$-4$p\pi$ dipole transition of Cu$^+$; the existence of a positive intensity, therefore, indicates the formation of Cu$^+$ (3$d^{10}$) sites in the sample due to annealing. By analyzing the intensity, we can evaluate $n_{\rm AN}$, taking the previous XAFS results of AS PCCO into account\cite{Asano2018a, Asano2018}. The value of $n_{\rm AN}$ obtained for LECO was 0.40 electrons per Cu, which is much larger than the electron density of 0.21 electrons per Cu in superconducting PCCO with $x$ = 0.16\cite{Asano2018}. Thus, a large number of electron carriers exist in superconducting LECO, while the number of doped electrons in non-superconducting NCO due to annealing is rather small ($n_{\rm AN}$ $\sim$ 0.05 electrons per Cu). Because the $\delta$ values of LECO and NCO are comparable with each other, the variation of the electronic state caused by annealing differs between the two compounds. We will discuss the doping process due to annealing for LECO and NCO later. 

Next, we analyzed the EXAFS spectra. Since the amount of removed oxygen is tiny, one may think that the change in the XANES spectra reflects only the local information around the removed oxygens. Thus, we examined that the variation of the electronic state due to annealing is caused throughout the LECO sample, from the structural point of view. 
Figure \ref{LECO_ChiR} shows $|\chi(r)|$, the absolute value of the Fourier transform of the EXAFS oscillations $k|\chi(k)|$ in the region $3 \leq k \leq 10~{\rm \AA}^{-1}$ for LECO and NCO. (The results for AS samples are shifted along the vertical direction by 0.4 ${\rm \AA}^{-2}$ for visual clarity.) 
In the figure, the Fourier-transform peaks/shoulders corresponding to Cu-O$_{\rm p}$, Cu-$RE$ ($RE$ = La, Eu, Nd), Cu-O$_{\rm RE}$, and Cu-Cu bonds can be seen. 
For both LECO and NCO, the overall shape of $|\chi(r)|$ is the same for AS and AN samples. Thus, the structural change due to annealing is negligible or small. 
However, the amplitude is smaller at each temperature for LECO than for NCO, meaning that there is considerably more static disorder in the superconducting LECO than in the non-superconducting NCO. 

The amplitude of each peak/shoulder decreases upon warming due to the thermal vibration of the atoms. The intensity of the peak at $\sim$1.5 $\AA$ has been analyzed to obtain the atomic displacement factor $C_2$ for the Cu-O$_{\rm p}$ bond. Figures \ref{DebyeWaller_v6}(a) and \ref{DebyeWaller_v6}(b) show the temperature dependence of $C_2$ for LECO and NCO, respectively. (To make different annealing effects on LECO and NCO more visible, $C_2$ for AS samples has been shifted along the vertical direction.) Upon warming, the $C_2$ of AN LECO increases more rapidly than that of AS LECO. This rapid increase indicates the softening of the Cu-O$_{\rm p}$ bond due to annealing, and is consistent with the result of the present XAFS; the large number of mobile carriers induced by annealing can soften phonons due to the screening effect. The softening of longitudinal optical phonons (corresponding to the Cu-O bond stretching mode) by electron doping has indeed been observed at around the $\Gamma$ point in T'-type cuprates\cite{DAstuto2002, Braden2005}. Therefore, we conclude that the variation of the electronic state due to annealing is a bulk phenomenon taking place throughout the entire LECO sample. On the other hand, no clear annealing effect on the thermal evolution of $C_2$ was observed in NCO, for which $n_{\rm AN}$ is rather small. 

	\begin{table}[tb]
 	\begin{center}
    \caption{Energy of Einstein oscillator $\hbar\omega_{\rm E}$ for as-sintered (AS) and annealed (AN) compounds of La$_{1.8}$Eu$_{0.2}$CuO$_{4}$ (LECO) and Nd$_{2}$CuO$_4$ (NCO), amplitude of softening due to annealing $\Delta{\omega_E}$, static atomic displacement factor $\sigma_{\rm s}$.}
    \begin{tabular}{ccccc} \hline
      ~~~~~~~~~~ & ~~~~~~~~ & ~$\hbar\omega_{\rm E}$ (meV)~ & ~$\Delta{\hbar\omega_E}$ (meV)~ & $\sigma_{\rm s}$ ($10^{-3}~{\rm \AA}^{2}$) \\ \hline
      LECO & AS & 44.3(4) & - & 2.0(1) \\ 
      LECO & AN & 40.6(4) & 3.7(4) & 4.1(1)\\ 
       NCO & AS
        & 44.8(5) & - & 0(1)\\ 
       NCO & AN
        & 44.2(4) & 0.6(5) & 0.5(1)\\ \hline
    \end{tabular}
    \label{Einstein_energy}
  	\end{center}
	\end{table}

	\begin{figure}[tb]
	\begin{center}
	\includegraphics[width=\linewidth]{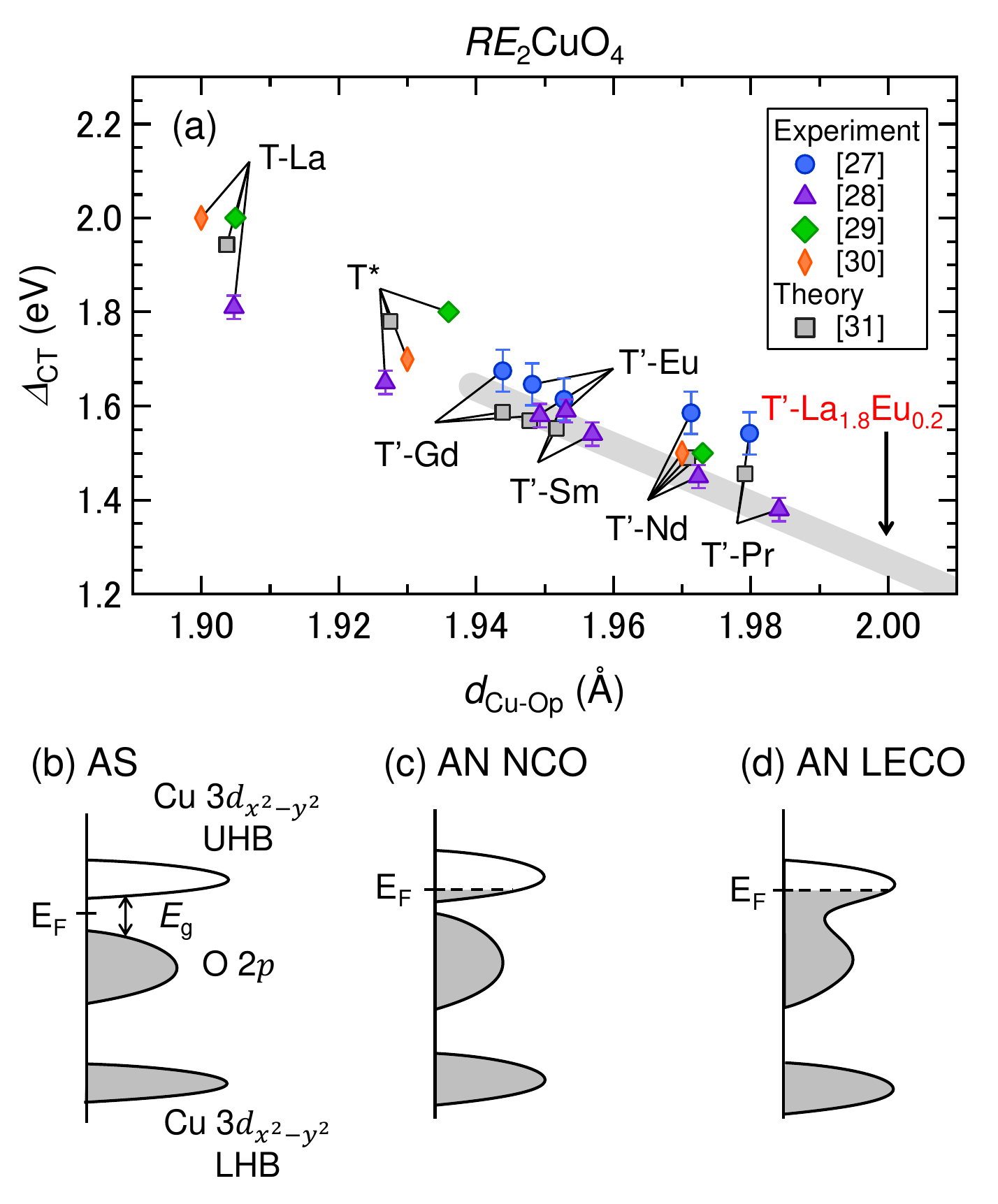}
	\caption{(Color online) (a) Energy of the charge transfer gap $\Delta_{\rm CT}$ as a function of the Cu-O$_{\rm p}$ bond-length $d_{\rm Cu-O_{\rm p}}$ for T-type La$_2$CuO$_4$ (T-La), T*-type cuprates (T*), T'-type Pr$_2$CuO$_4$ (T'-Pr), T'-type Nd$_2$CuO$_4$ (T'-Nd), T'-type Sm$_2$CuO$_4$ (T'-Sm), T'-type Eu$_2$CuO$_4$ (T'-Eu), T'-type Gd$_2$CuO$_4$ (T'-Gd), and  T'-type La$_{1.8}$Eu$_{0.2}$CuO$_4$ (T'-La$_{1.8}$Eu$_{0.2}$)\cite{Arima1991, Cooper1990, TokuraPRB1990, Tajima1990}. All of T'-type $RE_2$CuO$_4$ are as-sintered (AS) compounds.  $\Delta_{\rm CT}$ calculated by the ionic and cluster model is also plotted \cite{Ohta1991}. The gray solid line is a guide to the eye. Schematic picture of the density of states for (b) AS compounds, (c) annealed (AN) Nd$_{2}$CuO$_4$ (NCO) and (d) AN La$_{1.8}$Eu$_{0.2}$CuO$_4$ (LECO). See the text. }
	\label{CT_Cu-Obond_v5}
	\end{center}
	\end{figure}

Based on the Einstein model, the temperature dependence of $C_{2}$ was analyzed using the equation 
	\begin{eqnarray}
	C_{2} &=& \sigma_{\rm s} + ({\hbar}/2\mu\omega_E)^2{\rm coth}(\hbar{\omega_E}/2k_{\rm B}T), 
	\label{eq:debye_waller}
	\end{eqnarray}
where $\omega_E$ is the oxygen frequency of Einstein oscillators, and $\mu$ is the reduced mass of copper and oxygen atoms. Here,
$\hbar$ and $k_{\rm B}$ represent the reduced Planck constant and Boltzmann constant, respectively. 
For all samples, we assumed a temperature-independent $\sigma_{\rm s}$, since no evidence of structural transitions was observed in the measured temperature range. The values of $\hbar\omega_E$ evaluated for each sample are summarized in Table \ref{Einstein_energy}. In the AS compounds, the $\hbar\omega_E$ values for LECO and NCO are comparable with each other. However, $\omega_E$ in LECO decreases by about 10$\%$ due to annealing, while the annealing effect on $\omega_E$ in NCO is negligible. The quantitative analysis further revealed a larger $\sigma_{\rm s}$ in AN LECO than in AS LECO, indicating the enhancement of static disorder due to annealing.

\section{Discussion}
The present study has demonstrated for the first time a drastic variation of the electronic state in LECO associated with the appearance of superconductivity due to annealing. Here, we discuss the possible origin of this variation, as well as the reason that annealing has different effects on the electronic states in LECO and NCO. The focus of our discussion will be on the size of the CT gap. 

Mott insulators such as T-type La$_2$CuO$_4$ are characterized by the CT gap (with gap energy $\Delta_{\rm CT}$) between the Cu $3d_{x^{2}-y^{2}}$ UHB and the O 2$p$ band\cite{Zaanen1985} (Fig. \ref{CT_Cu-Obond_v5}(b)). Since a clear gap structure has been observed in the optical conductivity spectra of AS NCO\cite{Arima1991}, 
the similarity in the XANES spectra of AS LECO and AS NCO demonstrates that AS LECO is a Mott insulator. However, the anneal-induced electronic states of the two compounds are quite different. NCO’s $n_{\rm AN}$ (0.05 electrons per Cu) and $\delta$ (0.035) approximately follow the $n_{\rm AN}=2\delta$ relation. This means that the removal of one oxygen atom predominantly generates two electrons in the system, which is consistent with electron-doping into the UHB without disappearance of the CT gap, as shown in Fig. \ref{CT_Cu-Obond_v5}(c). By contrast, LECO’s $n_{\rm AN}$ of 0.40 electrons per Cu is far beyond the expected value from the $n_{\rm AN}=2\delta$ relation, given that $\delta$ $\leq$ 0.05. Thus, very many electrons exist in the superconducting sample. As was discussed in connection with the previous XANES results for PCCO, the emergence of so many electrons suggests the simultaneous generation of holes to maintain charge neutrality. This situation is difficult to explain by a picture based on electron doping into the UHB with finite $\Delta_{\rm CT}$. In the case of LECO, the removal of oxygen would cause the collapse of the CT gap, and as a result, both electrons and holes emerge at the Fermi level in the hybridized UHB and O 2$p$ bands. (See Fig. \ref{CT_Cu-Obond_v5}(d).) The collapse of the CT gap may occur by lowering the energy of UHB and/or broadening the bandwidth to touch on the O 2$p$ band. 

We speculate that such a distinct evolution is attributable to the different size of $\Delta_{\rm CT}$ in the AS compound. Figure \ref{CT_Cu-Obond_v5}(a) shows the $\Delta_{\rm CT}$ reported by optical studies \cite{Arima1991, Cooper1990, TokuraPRB1990, Tajima1990} as a function of the Cu-O bond-length $d_{\rm Cu-O}$ for T, T*, and T'-type $RE_2$CuO$_4$. $\Delta_{\rm CT}$, which is correlated with the Madelung energy at the Cu sites, is smaller in the compounds with lower coordination. Furthermore, in the T'-type $RE_2$CuO$_4$ with four coordination, $\Delta_{\rm CT}$ decreases with increasing $d_{\rm Cu-O}$. Since the $d_{\rm Cu-O}$ for AS LECO corresponding to $a$/2 is $2.00~{\rm \AA}$, LECO is expected to have the smallest $\Delta_{\rm CT}$ among $RE_2$CuO$_4$. A CT gap with small $\Delta_{\rm CT}$ easily collapses when the in-gap state is filled by a small amount of electron-doping and/or when the UHB is lowered by the removal of apical oxygen\cite{Adachi2013}. Therefore, even though the ground states of LECO and NCO are the same, reduction annealing induces a marked variation of the electronic state in LECO. 

Although the true ground state of T'-type $RE_2$CuO$_4$ cannot be elucidated due to the lack of information on the oxygen stoichiometry of the samples, the present study provides an important clue for understanding undoped superconductivity in T'-type cuprates. According to the results shown in Fig. \ref{CT_Cu-Obond_v5}(a), the CT gap potentially closes even in the AS compound with a sufficiently large value of $d_{\rm Cu-O}$, and therefore, the metallic state will appear. Combined with the experimental fact that superconductivity is induced with a smaller Ce concentration in T'-type $RE_{2-x}$Ce$_x$CuO$_4$ having a larger in-plane lattice constant\cite{Naito2002, Fujita2003, Krockenberger2008, Krockenberger2014}, this implies that undoped superconductivity could be realized in T'-type $RE_2$CuO$_4$. Thus, the ground state of T'-type $RE_{2-x}$Ce$_x$CuO$_4$ can be controlled by the Cu-O bond-length. This opens the way for a unified understanding of the physical properties of T'-type cuprates from the viewpoint of the size of $\Delta_{\rm CT}$. 

Finally, we will briefly discuss static disorder in T'-type $RE_2$CuO$_4$. In the $RE_2$CuO$_4$ cuprate oxides, the T'-type structure tends to transform into the T-type one as the average radius $R_{\rm AV}$ of the rare-earth ion increases \cite{Xiao1989, Chaudhri1997, Ikeda2002, Zhang2002, Imai2007}. This is to alleviate the size mismatch between Cu-O-Cu and $RE$-$RE$ bonds. Our EXAFS measurements revealed a larger value of $\sigma_{\rm s}$ in LECO ($R_{\rm AV}$ = 1.164 $\AA$) than in NCO ($R_{\rm AV}$ = 1.123 $\AA$). 
This is probably due to LECO’s greater structural instability, since LECO locates near the phase boundary between T' and T structures against $R_{\rm AV}$. In the vicinity of the phase boundary, the elongation of the in-plane lattice constant due to annealing easily enhances the structural instability and increases $\sigma_{\rm s}$, as seen in Table \ref{Einstein_energy}. Muon spin relaxation measurements suggest the coexistence of short-range magnetic order with superconductivity in the ground state of AN LECO\cite{Adachi2016}. From the structural point of view, the existence of a magnetic order in superconducting LECO is attributable to a partial localization of carriers due to the large structural disorder. That is, the structural disorder accompanied by the disorder of the electrostatic potential causes the localization of carriers, leading to the appearance of static magnetism in the competing superconducting state. Thus, $T_{\rm c}$ in LECO could be increased by the suppression of magnetic order through the relaxation of structural disorder.

In summary, reduction annealing effects on the electronic states at Cu sites in LECO and NCO have been investigated by Cu $K$-edge XAFS measurements. The analysis of XANES spectra has revealed a significant evolution of the electronic state in LECO due to annealing with a large induced electron density $n_{\rm AN}$ of 0.40 electrons per Cu. For NCO, the variation is much smaller, with $n_{\rm AN}$ only 0.05 electrons per Cu. The EXAFS analysis showed evidence of phonon softening of the Cu-O bond in the CuO$_2$ plane due to annealing for LECO but not for NCO. Therefore, the electronic states for the two compounds vary by distinct processes, although $\delta$, the amount of oxygen loss due to annealing, is comparable for the two systems. The distinct evolution of the electronic states for LECO and NCO can be attributed to the difference in size of the charge-transfer energy gap in the AS compounds; the electronic state in LECO, which has a smaller $\Delta_{\rm CT}$, is more sensitive to the oxygen removal that could introduce the electron-doping and/or variation of the band structure.

\section*{Acknowledgements}
We thank Y. Kimura for support in the analysis of XAFS data. The synchrotron radiation experiments were performed at the BL01B1 and BL14B1 of SPring-8 with the approval of the Japan Synchrotron Radiation Research Institute (JASRI) (Proposal No. 2016A1603 and No. 2017B3611).  M. F. is supported by Grant-in-Aid for Scientific Research (A) (16H02125), K. I. is supported by Grant-in-Aid for Scientific Research (B) (16H04004), and Y.K. is supported by Grant-in-Aid for Scientific Research (B) (17H02915). 

\bibliographystyle{aps}
\bibliography{XAFS_LECO.bib}

\begin{thebibliography}{10}
\providecommand{\url}[1]{\texttt{#1}}
\providecommand{\urlprefix}{URL }
\providecommand{\eprint}[2][]{\url{#2}}

\bibitem{Matsumoto2009}
O.~Matsumoto, A.~Utsuki, A.~Tsukada, H.~Yamamoto, T.~Manabe, and M.~Naito,
  Physica C \textbf{469}, 924 (2009).

\bibitem{Asai2011}
S.~Asai, S.~Ueda, and M.~Naito, Physica C \textbf{471}, 682 (2011).

\bibitem{Takamatsu2012}
T.~Takamatsu, M.~Kato, T.~Noji, and Y.~Koike, Appl. Phys. Express \textbf{5},
  073101 (2012).

\bibitem{Das2009}
H.~Das and T.~Saha-Dasgupta, Phys. Rev. B \textbf{79}, 134522 (2009).

\bibitem{Weber2010a}
C.~Weber, K.~Haule, and G.~Kotliar, Nat. Phys. \textbf{6}, 574 (2010).

\bibitem{Weber2010}
C.~Weber, K.~Haule, and G.~Kotliar, Phys. Rev. B \textbf{82}, 125107 (2010).

\bibitem{Takagi1989}
H.~Takagi, S.~Uchida, and Y.~Tokura, Phys. Rev. Lett. \textbf{62}, 1197 (1989).

\bibitem{Brinkmann1995}
M.~Brinkmann, T.~Rex, H.~Bach, and K.~Westerholt, Phys. Rev. Lett. \textbf{74},
  4927 (1995).

\bibitem{Fujita2003}
M.~Fujita, T.~Kubo, S.~Kuroshima, T.~Uefuji, K.~Kawashima, K.~Yamada,
  I.~Watanabe, and K.~Nagamine, Phys. Rev. B \textbf{67}, 014514 (2003).

\bibitem{Horio2015a}
M.~Horio, T.~Adachi, Y.~Mori, A.~Takahashi, T.~Yoshida, H.~Suzuki, K.~Okazaki,
  K.~Ono, H.~Kumigashira, M.~Arita, H.~Namatame, M.~Taniguchi, D.~Ootsuki,
  M.~Takahashi, T.~Mizokawa, Y.~Koike, and A.~Fujimori, Nat. Commun.
  \textbf{7}, 10567 (2015).

\bibitem{Wei2016}
H.~I. Wei, C.~Adamo, E.~A. Nowadnick, E.~B. Lochocki, S.~Chatterjee, J.~P. Ruf,
  M.~R. Beasley, D.~G. Schlom, and K.~M. Shen, Phys. Rev. Lett. \textbf{117},
  147002 (2016).

\bibitem{Song2017}
D.~Song, G.~Han, W.~Kyung, J.~Seo, S.~Cho, B.~S. Kim, M.~Arita, K.~Shimada,
  H.~Namatame, M.~Taniguchi, Y.~Yoshida, H.~Eisaki, S.~R. Park, and C.~Kim,
  Phys. Rev. Lett. \textbf{118}, 137001 (2017).

\bibitem{Horio2018e}
M.~Horio, Y.~Krockenberger, K.~Koshiishi, S.~Nakata, K.~Hagiwara, M.~Kobayashi,
  K.~Horiba, H.~Kumigashira, H.~Irie, H.~Yamamoto, and A.~Fujimori, Phys. Rev.
  B \textbf{98}, 020505(R) (2018).

\bibitem{Horio2018b}
M.~Horio, Y.~Krockenberger, K.~Yamamoto, Y.~Yokoyama, K.~Takubo, Y.~Hirata,
  S.~Sakamoto, K.~Koshiishi, A.~Yasui, E.~Ikenaga, S.~Shin, H.~Yamamoto,
  H.~Wadati, and A.~Fujimori, Phys. Rev. Lett. \textbf{120}, 257001 (2018).

\bibitem{Horio2018f}
M.~Horio and A.~Fujimori, J. Phys. Condens. Matter \textbf{30}, 503001 (2018).

\bibitem{Yokoyama2006}
H.~Yokoyama, M.~Ogata, and Y.~Tanaka, J. Phys. Soc. Jpn. \textbf{75}, 114706
  (2006).

\bibitem{Adachi2013}
T.~Adachi, Y.~Mori, A.~Takahashi, M.~Kato, T.~Nishizaki, T.~Sasaki,
  N.~Kobayashi, and Y.~Koike, J. Phys. Soc. Jpn. \textbf{82}, 063713 (2013).

\bibitem{Yamazaki2017}
K.~Yamazaki, T.~Yoshioka, H.~Tsuchiura, and M.~Ogata, J. Phys. Conf. Ser.
  \textbf{871}, 012009 (2017).

\bibitem{Asano2018}
S.~Asano, K.~Ishii, D.~Matsumura, T.~Tsuji, T.~Ina, K.~M. Suzuki, and
  M.~Fujita, J. Phys. Soc. Jpn. \textbf{87}, 094710 (2018).
  
 \bibitem{Sato}
K.~Sato, S.~Torii, and
  M.~Fujita, unpublished data.

\bibitem{Suzuki2019}
K.~M. Suzuki, S.~Asano, H.~Okabe, A.~Koda, R.~Kadono, I.~Watanabe, and
  M.~Fujita, arXiv:1901.11233  (2019).

\bibitem{Liang1995}
G.~Liang, Y.~Guo, D.~Badresingh, W.~Xu, Y.~Tang, M.~Croft, J.~Chen, A.~Sahiner,
  B.-h. O, and J.~T. Markert, Phys. Rev. B \textbf{51}, 1258 (1995).

\bibitem{Oyanagi1990}
H.~Oyanagi, Y.~Yokoyama, H.~Yamaguchi, Y.~Kuwahara, T.~Katayama, and
  Y.~Nishihara, Phys. Rev. B \textbf{42}, 10136 (1990).

\bibitem{Kosugi1990}
N.~Kosugi, Y.~Tokura, H.~Takagi, and S.~Uchida, Phys. Rev. B \textbf{41}, 131
  (1990).

\bibitem{Asano2018a}
S.~Asano, K.~M. Suzuki, D.~Matsumura, K.~Ishii, T.~Ina, and M.~Fujita, J. Phys.
  Conf. Ser. \textbf{969}, 012051 (2018).

\bibitem{DAstuto2002}
M.~D'Astuto, P.~K. Mang, P.~Giura, A.~Shukla, P.~Ghigna, A.~Mirone, M.~Braden,
  M.~Greven, M.~Krisch, and F.~Sette, Phys. Rev. Lett. \textbf{88}, 167002
  (2002).

\bibitem{Braden2005}
M.~Braden, L.~Pintschovius, T.~Uefuji, and K.~Yamada, Phys. Rev. B \textbf{72},
  184517 (2005).

\bibitem{Arima1991}
T.~Arima, K.~Kikuchi, M.~Kasuya, S.~Koshihara, Y.~Tokura, T.~Ido, and
  S.~Uchida, Phys. Rev. B \textbf{44}, 917 (1991).

\bibitem{Cooper1990}
S.~L. Cooper, G.~A. Thomas, A.~J. Millis, P.~E. Sulewski, J.~Orenstein, D.~H.
  Rapkine, S.-W. Cheong, and P.~L. Trevor, Phys. Rev. B \textbf{42}, 10785
  (1990).

\bibitem{TokuraPRB1990}
Y.~Tokura, S.~Koshihara, T.~Arima, H.~Takagi, S.~Ishibashi, T.~Ido, and
  S.~Uchida, Phys. Rev. B \textbf{41}, 11657 (1990).

\bibitem{Tajima1990}
S.~Tajima, S.~Uchida, S.~Ishibashi, T.~Ido, H.~Takagi, T.~Arima, and Y.~Tokura,
  Physica C \textbf{168}, 117 (1990).

\bibitem{Ohta1991}
Y.~Ohta, T.~Tohyama, and S.~Maekawa, Phys. Rev. Lett. \textbf{66}, 1228 (1991).

\bibitem{Zaanen1985}
J.~Zaanen, G.~A. Sawatzky, and J.~W. Allen, Phys. Rev. Lett. \textbf{55}, 418
  (1985).

\bibitem{Naito2002}
M.~Naito, S.~Karimoto, and A.~Tsukada, Supercond. Sci. Technol. \textbf{15},
  1663 (2002).

\bibitem{Krockenberger2008}
Y.~Krockenberger, J.~Kurian, A.~Winkler, A.~Tsukada, M.~Naito, and L.~Alff,
  Phys. Rev. B \textbf{77}, 060505(R) (2008).

\bibitem{Krockenberger2014}
Y.~Krockenberger, B.~Eleazer, H.~Irie, and H.~Yamamoto, J. Phys. Soc. Jpn.
  \textbf{83}, 114602 (2014).

\bibitem{Xiao1989}
G.~Xiao, M.~Z. Cieplak, and C.~L. Chien, Phys. Rev. B \textbf{40}, 4538 (1989).

\bibitem{Chaudhri1997}
M.~Chaudhri, K.~B. Modi, K.~Jadhav, and G.~Bichile, Pramana \textbf{48}, 883
  (1997).

\bibitem{Ikeda2002}
Y.~Ikeda, K.~Yamada, Y.~Kusano, and J.~Takada, Physica C \textbf{378--381}, 395
  (2002).

\bibitem{Zhang2002}
C.~Zhang and Y.~Zhang, Journal of Physics: Condensed Matter \textbf{14}, 7383
  (2002).

\bibitem{Imai2007}
Y.~Imai, M.~Kato, Y.~Takarabe, T.~Noji, T.~Adachi, and Y.~Koike, Physica C
  \textbf{460--462}, 395 (2007).

\bibitem{Adachi2016}
T.~Adachi, A.~Takahashi, K.~M. Suzuki, M.~A. Baqiya, T.~Konno, T.~Takamatsu,
  M.~Kato, I.~Watanabe, A.~Koda, M.~Miyazaki, R.~Kadono, and Y.~Koike, J. Phys.
  Soc. Jpn. \textbf{85}, 114716 (2016).

\end{thebibliography}

\end{document}